\documentstyle[11pt,aaspp4]{article}
\input epsf.sty

\def\Deg{\hbox{${}^\circ$\llap{.}}}
\def\Min{\hbox{${}^{\prime}$\llap{.}}}
\def\Sec{\hbox{${}^{\prime\prime}$\llap{.}}}

\def\numceph{33\thinspace}
\def\numcephall{49\thinspace}
\def\lowperiod{8\thinspace}
\def\upperperiod{47\thinspace}
\def\distmod0{30.78\thinspace}
\def\distmoderrorrandom{0.14\thinspace}
\def\distmoderrorsys{0.10\thinspace}
\def\distmodv{30.93\thinspace}
\def\distmodi{30.87\thinspace}
\def\distmodverr{0.06\thinspace}
\def\distmodierr{0.05\thinspace}

\received{}
\lefthead{Sakai {\it et al.}}
\righthead{Cepheid distance to NGC 3319}

\begin{document}

\begin{center}
{\bf The Hubble Space Telescope Extragalactic Distance Scale
Key Project XXIII. \\
The Discovery of Cepheids In NGC 3319\footnote
{Based on observations with the NASA/ESA {\it Hubble Space Telescope}, obtained at the Space Telescope Science Institute, operated by AURA, Inc. under NASA contract No. NAS5-26555.}}

\medskip
\bigskip

Shoko Sakai$^{2}$, 
Laura Ferrarese$^{3,4}$, 
Robert C. Kennicutt$^{5}$, 
John A. Graham$^{6}$,
N. A. Silbermann$^{7}$
Jeremy R. Mould$^{8}$, 
Wendy L. Freedman$^{9}$,
Fabio Bresolin$^{10}$
Holland C. Ford$^{11}$, 
Brad K. Gibson$^{12}$,
Mingsheng Han$^{13}$, 
Paul Harding$^{5}$,
John G. Hoessel$^{14}$,
John P. Huchra$^{15}$, 
Shaun M. Hughes$^{16}$, 
Garth D. Illingworth$^{17}$,
Daniel Kelson$^{6}$,
Lucas Macri$^{15}$,
Barry F. Madore$^{7}$, 
Randy L. Phelps$^{9}$, 
Abhijit Saha$^{2}$,
Kim M. Sebo$^{8}$,
Peter B. Stetson$^{18}$,
Anne Turner$^{5}$

\end{center}

\altaffiltext{2}{Kitt Peak National Observatory, National Optical Astronomy Observatories, Tucson AZ 85726, USA}

\altaffiltext{3}{California Institute of Technology, Pasadena CA 91125, USA}

\altaffiltext{4}{Hubble Fellow}

\altaffiltext{5}{Steward Observatories, University of Arizona, Tucson AZ 85721, USA}

\altaffiltext{6}{Department of Terrestrial Magnetism, Carnegie Institution of 
Washington, Washington, DC 20015, USA}

\altaffiltext{7}{Infrared Processing and Analysis Center, Jet Propulsion Laboratory, 
California Institute of Technology, Pasadena CA 91125, USA}

\altaffiltext{8}{Mount Stromlo and Siding Spring Observatories, 
Australian National University, Weston Creek, 
ACT 2611, Australia}

\altaffiltext{9}{Carnegie Observatories, Pasadena CA 91101, USA}

\altaffiltext{10}{European Southern Observatory, Karl--Schwarzschild--Strasse 2, 
D--85748 Garching, Germany}

\altaffiltext{11}{Space Telescope Science Institute, Baltimore MD 21218, USA}
\altaffiltext{12}{Center for Astrophysics \& Space Astronomy, Department of Astrophysical \&
Planetary Sciences, University of Colorado, Campus Box 389, Boulder, CO, USA 80309-0389}

\altaffiltext{13}{Avanti Corp,  46871 Bayside Parkway, Fremont, CA 94538, USA}

\altaffiltext{14}{Department of Astronomy, University of Wisconsin, Madison, WI, 53706, USA}

\altaffiltext{15}{Harvard Smithsonian Center for Astrophysics, Cambridge, MA 02138, USA}

\altaffiltext{16}{Institute of Astronomy, Madingley Road, Cambridge, UK CB3 0HA}

\altaffiltext{17}{Lick Observatory, University of California, Santa Cruz CA 95064, USA}

\altaffiltext{18}{Dominion Astrophysical Observatory, Herzberg Institute of Astrophysics,
National Research Council, 5071 West Saanich Rd., Victoria, BC V8X 4M6, Canada}

\begin{center}
Accepted for Publication in {\it Astrophysical Journal}
\end{center}
\bigskip
\bigskip

\begin{abstract}

The distance to NGC~3319 has been determined from 
Cepheid variable stars as part of the Hubble Space Telescope
Key Project on the Extragalactic Distance Scale.
Thirteen and four epochs of observations, using filters F555W (V) and 
F814W (I) respectively, were made with the Wide Field Planetary Camera~2.
Thirty--three Cepheid variables between periods of 8 and 47 days
were discovered.
Adopting a Large Magellanic Cloud distance modulus of 18.50 
$\pm$ 0.10 mag and extinction of E(V-I)=0.13 mag, a true 
reddening-corrected distance modulus (based on an analysis 
employing the ALLFRAME software package) of 
\distmod0 $\pm$ \distmoderrorrandom (random) $\pm$ \distmoderrorsys 
(systematic)  mag and the extinction of
$E(V-I) = 0.06$ mag were determined for NGC~3319.
This galaxy is the last galaxy observed for the HST H$_0$ Key Project.

\end{abstract}
\section{Introduction}

The main objective of the {\it Hubble Space Telescope Key Project
on Extragalactic Distance Scale} (hereafter H$_0$ Key Project) 
is to measure the value of the Hubble
constant (H$_0$) to an accuracy of 10\%. This is achieved by
searching for and then measuring the periods and magnitudes of 
Cepheid variable stars in 18 nearby spiral galaxies.  
The accurate Cepheid distances of these galaxies are then used
to calibrate several secondary distance indicators, including
the Tully--Fisher relation for late--type spiral galaxies,
surface brightness fluctuation method, planetary nebulae luminosity 
function, and Type Ia supernovae, which 
enable us to measure the Hubble constant at distances
where the peculiar velocities are significantly small compared
to their redshifts.

NGC~3319 is the last galaxy to be observed for this project.
Distances to seventeen other galaxies targeted by the H$_0$
Key Project have been published in the Papers I through XXII
of this series.
NGC~3319 is an SB(rs)cd galaxy with a prominent bar, located at 
R.A. (J2000) = 10$^h$ 39$^m$ 09.8$^s$ and  DEC (J2000) = 
41$^d$ 41$^m$ 15.9$^s$ at a heliocentric radial velocity of 742 $\pm$ 10
km s$^{-1}$ (Tully 1988).
Although Moore \& Gottesman (1998) pointed out that the HI distribution is
asymmetric, NGC~3319 shows no sign of interaction.
It is a fairly isolated galaxy with its closest galaxy
in the velocity range of 200 and 1200 km s$^{-1}$ almost 4$^{\circ}$ away.
Bottinelli et al. (1993) 
obtained $i=56$ using an RC2 axial ratio, making it an ideal Tully-Fisher
calibrator.

\section{Observations and Data Reduction}

\subsection{Observations}

The observing strategy of H$_0$ Key Project galaxies has already been explained
in detail by previous papers in this series (e.g. Kennicutt, Freedman 
\& Mould 1995), thus we will review here only the observations 
pertaining specifically to the NGC~3319 data set.

A ground--based image of NGC~3319, obtained using the 1.2m telescope
at Fred L. Whipple Observatory, is shown in Figure~1,
with the region observed by the HST Wide Field Planetary Camera~2
(WFPC2) indicated by a footprint.
WFPC2 is comprised of four $800 \times 800$ CCDs.
The Planetary Camera (PC) chip covers the smallest field of view,
37 arcsec on the side.  Three Wide Field Camera (WFC)
chips have $80 \times 80$ arcsecond fields of view each.
We refer to these three chips as Chips 2, 3 and 4,
clockwise from the PC in Figure~1.

$HST$ WFPC2 observations of NGC~3319 were carried out over a two--month
period (November 11, 1997 -- January 3rd, 1998), with a single
epoch previsit on January 1st, 1996.
The previsit epoch, almost two years before the start of the main observing
campaign, was particularly helpful for constraining the periods of
long--period Cepheid variables.
A total of thirteen epochs of F555W (V) and 
four epochs of F814W (I) cosmic--ray split observations were made.
The previsit observation was done using only the F555W filter.
A summary of dates of observations, exposure times and filters used for
each epoch is given in Table~1.
The exposure times are 1100 and 1300 seconds for F555W and F814W filters
respectively, except for epoch \#7 where only a single 2200s
exposure in F555W was taken.

\subsection{Data Reduction: ALLFRAME}

The NGC~3319 HST/WFPC2 data were first calibrated using a standard
routine at Space Telescope Science Institute (STScI) which included
the following procedures (see Holtzman et al. 1995 for details): 
correction of A/D errors, subtraction of a bias level and subsequently 
of a superbias frame, subtraction of a dark frame, and then 
a division by a flat field frame.
Furthermore, we performed several additional procedures:
(1) bad pixels on the images were masked using the data quality
files provided by the $HST$ data--processing pipelines;
(2) the vignetted edges were blocked by applying the vignetting
masks; (3) the geometric distortion of the WFPC2 optics was
corrected using pixel area maps; and finally (4) each frame was
multiplied by four and converted to short integers, thereby
reducing the disk space usage.

All of the photometric reduction of NGC~3319 HST data were carried
out using the DAOPHOT/ALLFRAME family of software (Stetson 1994).  
The detailed description of reduction procedures using ALLFRAME
can be found in Stetson et al. (1998), thus we will provide, again,
only a brief summary of NGC~3319 data reduction processes below.

Instead of applying cosmic--ray corrections to the individual
galaxy frames, all 33 images were median--combined
to create a cosmic--ray free, clean master frame.
The master star list was created using the automated star-finding routines
from the DAOPHOT and ALLSTAR packages.
This list was fed into ALLFRAME and used to extract
profile--fitting stellar photometry from 33 individual frames,
using point spread functions (PSFs) constructed from $HST$ WFPC2 observations
of the globular clusters Pal 4 and NGC~2419.

The procedures to convert F555W and F814W instrumental magnitudes
to the calibrated Landolt (1992) system are described in detail in 
Hill et al. (1998).  
Briefly, the instrumental magnitudes were first brought to the
Holtzman et al. (1995) 0\Sec5--arcsec aperture system.
We determined the aperture--correction
to transform the ALLFRAME magnitudes to a 0\Sec5--arcsec aperture as follows.
First, 30 to 40 isolated, relatively bright stars were selected on each chip.
Then all the stars, except those isolated ones, were subtracted from the
original image, leaving a clean image suitable for good sky estimates.
Aperture magnitudes at 12 different radii, ranging from 0\Sec15 to 0\Sec50,
were measured for these selected stars.  
Using a growth--curve analysis provided by 
DAOGROW (Stetson 1990), the 0\Sec5
aperture magnitude for each star was determined, then
compared with the corresponding PSF magnitude to calculate the 
aperture correction.
For each chip, the aperture correction
was determined by taking the weighted mean of the aperture corrections
of all the selected isolated stars, whose values typically ranged
between $-0.10$ and $+0.05$ mag.

The transformation equation to convert the 0\Sec5--aperture magnitudes, $m$,
to the Hill et al. (1997) system, $M$, is expressed as:
\begin{equation}
M = m + 2.5\log t + C1 + C2 * (V-I) + C3 * (V-I)^2 + AC,
\end{equation}
where $t$ is the exposure time, C1 through C3 are constants and 
A.C. is the aperture correction.
The coefficients C2 and C3 for the color--corrections were adopted
from Holtzman et al. (1995) and are the same for all four chips.
The constant C1 consists of several terms, such as the long--exposure
zero points and a pixel area map normalization correction.
Table 2 lists the coefficients for each chip.
A full discussion of all the terms and their magnitudes can be
found in Hill et al. (1998).

In Table~3, we list the V and I magnitudes
of 15--20 selected bright, isolated reference stars for each chip.
Columns are: (1) Star ID, (2) Chip numbers, (3) \& (4)
The $(x,y)$ coordinates, referring to the pixel
positions of stars on the first image of the previsit observations (U2S75501T),
(5) \& (6) RA and DEC (J2000), (7) V magnitude, and (8) I magnitude.
Unlike other galaxies in the H$_0$ Key Project, 
only the ALLFRAME photometry analysis was undertaken for NGC~3319.
However, as an external zero--point check, we used the DoPHOT
package to obtain the photometry of stars on eight F555W (epochs \#2, 3, 11 and
12), and six (epochs \#2, 3 and 12) F814W images.  
The ALLFRAME and DoPHOT photometry results agree within 0.1 mag,
which is roughly 1.5 times the $HST$ photometric zero--point errors.

\section{Cepheid Identification}

Cepheid variable stars were searched for using 
a template light--curve fitting procedure, called TRIAL (Stetson 1996).  
A candidate variable star was retained as a possible Cepheid
variable only if the following criteria were met:
(1) high--quality photometric data (photometry error $\leq$ 0.3 mag)
were obtained for at least 10 epochs in V;
(2) the shape of the light curve showed a short rise time followed
by a longer decline, which is characteristic of a Cepheid light curve;
and (3) its V--I color is consistent with that of the instability strip.

The TRIAL algorithm is based on Welch \& Stetson's method (1993),
which is based on the idea that residuals from intrinsic variability
are strongly correlated while photometric errors are essentially random.
We set the variability index low (0.5) at first,
so as to make the Cepheid candidate list as complete as possible.
This resulted in a list of 571 variable star candidates.
Their individual light curves were then visually inspected to select
those stars that would meet all three criteria listed above.
In addition, we have excluded those candidate Cepheid variables
that are in a very crowded region.
As a result, we found a total of \numceph ``high--quality''
Cepheid variable stars which will be used in the following
sections to derive the distance to NGC~3319.
In Table 4, the Cepheid ID number, 
the $(x,y)$ coordinates on the reference
frame of the U2S75501T dataset, and RA and DEC (Epoch 2000) are summarized.
Epoch by epoch V and I magnitudes are listed in Tables 5 and 6 respectively.

\subsection{Mean Magnitudes}

Following the standard procedure adopted by the H$_0$ Key Project,
we measured the mean magnitudes of the Cepheid variable stars
using two different definitions: (1) intensity--weighted
mean magnitude:
\begin{equation}
m = -2.5 \log \sum_{i=1}^n \frac{1}{n} 10^{-0.4 m_i},
\end{equation}
and (2) phase--weighted mean magnitude (Saha \& Hoessel 1990):
\begin{equation}
m = -2.5 \log \sum_{i=1}^n 0.5(\phi_{i+1} - \phi_{i-1}) 10^{-0.4 m_i},
\end{equation}
where $n$ is the total number of observations, and $m_i$
and $\phi_i$ are the magnitude and phase of the $i$-th observation
respectively, in order of increasing phase.

Freedman (1988) showed that there was a good correspondence
between the V and I light curves and that the V-band light curve
can be mapped onto the I--light curve by a simple scaling by a ratio 0.51
of their amplitudes.  This ratio has an error of $\sim$0.04 mag if a
single I--band observation was used to determine the mean I magnitude.
Thus for the case of NGC~3319 where four I--band observations were made,
the error associated with this scaling should be negligible.
Using this empirical fact, the I--band mean magnitudes are corrected as follows.
First, the difference between the mean V magnitude from the complete
V dataset and a mean using only those V data points in common with the
I observations (four epochs) is calculated.
This difference is multiplied by a factor 0.51, which is then added
to the phase--weighted mean I magnitude (which was originally determined
from the data from only four epochs).
This correction is in most cases very small, on the order of $0.01 - 0.05$ mag.
In a couple of cases of poorly sampled I light curves, the correction was as large
as 0.08 mag (see Table 7).

The intensity--weighted and phase--weighted mean magnitudes (both V and I),
and periods for all \numceph Cepheid variable stars are summarized in Table 7.
It also lists the additive correction that was applied to 
the I--band phase--weighted magnitudes to compensate for their limited
phase coverage.
For uniform sampling, the phase--weighted and intensity--weighted
mean magnitudes are the same.  For the case of NGC~3319 observations,
the two mean magnitudes agree within $-0.014 \pm 0.05$ mag.
In measuring mean magnitudes, these observations with corresponding
errors exceeding 0.3 mag were excluded (see Tables 5 and 6).
In Table 5, also listed are whether a Cepheid variable was included
in the NGC~3319 distance estimate. This is indicated by ``y'' or ``n''
in the third column.
We excluded C33, whose period is only 8 days, from the following
analyses, since short--period Cepheids are often over--tone pulsators.  
C01 was also excluded. Its period is estimated to be longer than 47 days.
However, the photometric data for this Cepheid does not exist
for the pre--visit observations as its position was unfortunately
outside the boundary of WF4 chip for this epoch.  Thus, the period 
of the Cepheid variable C01 is too long to be well determined
by the 1997/98 data.
C12 was also excluded from the further distance determination
as this Cepheid variable is located too close to the edge of the WF3 chip.
Furthermore, C14 and C29 were not included in the final sample either
since the former is a possible blend, and the latter is near a very bright
cluster.  Although C11, C25 and C31 appear to be in crowded regions, 
because the star--subtraction near these stars are good, and they are not
obvious blends, we have included them in the distance determination.

Figures 2 and 3 show the finding charts for
all the high--quality Cepheid variables found in NGC~3319,
for individual chip and Cepheid, respectively.
The V-- and I--band light curves for each Cepheid are shown in Figure 4.
The magnitudes plotted in Figure 4 are usually the means
of the cosmic--ray split images.  
However there are exceptions, such as for epoch 7 where there was
only one exposure taken, and whenever an ALLFRAME magnitude error
exceeded 0.3 mag, in which case these magnitudes were discarded as being
likely to have been affected by a cosmic ray.
In Figure 5, we show an I versus (V--I) color--magnitude
diagram of all the stars found in NGC~3319.  The Cepheid variables,
indicated by larger solid circles, lie in the instability strip, as expected.
In Figure~6, the V and I--band period--luminosity relations are shown.
The solid lines indicate least squares fits while the dotted lines
represent their 2$\sigma$ dispersion from the LMC PL--relations.
In all the Figures (2 through 6), phase--weighted magnitudes were plotted.

\section{The Cepheid Distance to NGC~3319}

Following other papers in this series, the apparent V and I distance
moduli to NGC~3319 are estimated based on the absolute calibrations
adopted from Madore \& Freedman (1991) which are based on a consistent set of
32 Large Magellanic Cloud (LMC) Cepheid variables with $BVI$ 
observations.  The V and I relations are expressed as:
\begin{equation}
M_V  = -2.76 (\log_{10}P - 1.0) - 4.16
\end{equation}
\begin{equation}
M_I = -3.06 (\log_{10}P - 1.0) - 4.87
\end{equation} 
which assume an LMC true modulus of $\mu_{LMC} = 18.50 \pm 0.10$ 
and  average line-of-sight reddening of  $E(B-V) =  0.10$ mag.
The apparent distance modulus for the NGC~3319 data at each wavelength was 
determined by minimizing the {\it rms} deviations of the observed data about 
the ridge line, with the slopes fixed to those given by the above equations.
This minimizes the bias due to the incompleteness at short periods in
the NGC~3319 Cepheid sample.

Using the phase--weighted magnitudes,
V and I apparent moduli for NGC~3319 are estimated to be
$\mu_V$ = \distmodv $\pm$ \distmodverr and $\mu_I$ = 
\distmodi $\pm$ \distmodierr.
The true distance modulus is determined through a relation:
$\mu_0 = \mu_V - A_V = \mu_I - A_I$.
Adopting $A_V/E(V-I) = 2.45$ and $R_V = A_V/E(B-V) = 3.3$, which are 
consistent with Dean, Warren \& Cousins (1978), Cardelli, Clayton \& 
Mathis (1989) and Stanek (1996), we derive the true distance modulus
of NGC~3319 of:
\begin{equation}
(m-M)_0 = \distmod0 \pm \distmoderrorrandom \mbox{(random)} \pm 
\distmoderrorsys \mbox{(systematic)}.
\end{equation}
If the intensity--weighted magnitudes were used instead,
the true distance modulus would have been $30.79 \pm 0.09$ mag,
which agrees with the phase--weighted distance modulus.

Errors quoted above are comprised of several independent sources
(Table 8):
(1) uncertainties in the LMC $PL$ relation zero points and also
in the adopted distance to the LMC itself.  
For the error in the LMC
true modulus, we adopt $\pm$0.10 mag (Westerlund 1997);
(2) uncertainties in the extinction corrections, specifically those
in the adopted value of $R_V$;
(3) uncertainties in the metallicity dependence of the $PL$ relation,
which could affect the distance to NGC~3319 as we have assumed the
same metallicities for both galaxies, NGC~3319 and the LMC;
(4) uncertainties arising from the difference in the values of
$R_V$ for NGC~3319 and LMC;
(5) uncertainties in photometry, in particular the error in the zero point
of the $HST$ calibration, and also the difference in the zero points
of ALLFRAME and DoPHOT photometry; and,
(6) uncertainties in fitting the LMC $PL$ relations to the NGC~3319 dataset.

Most of the errors are discussed in detail in the previous papers of
this series (cf. Ferrarese et al. 1998, Hughes et al. 1998).
Here, we shall comment only on the uncertainty in the NGC~3319
distance due to the metallicity dependence of the Cepheids' PL relation.
Kennicutt et al. (1998) measured the distance to M101
from the observations of two fields in this spiral galaxy.
Attributing the source of the difference in the two distance moduli to the
metallicity difference in two fields, they obtained the Cepheid $PL$
relation metallicity dependence of $-0.24 \pm 0.16$ mag/dex.
The mean difference in $[O/H]$ between the WFPC2 field in NGC~3319 and the LMC
is $\Delta [O/H] = -0.1 \pm 0.15$ (in the sense that NGC~3319 is less metal rich) 
(Zaritsky, Kennicutt \& Huchra, 1994).
Due to the uncertainties attributed to the metallicity correction 
of Kennicutt et al. (1998), and to be consistent with previous
papers in this series, we will not apply the metallicity correction,
but merely state this correction value as a potential systematic error
(0.24 mag dex$^{-1}$ $\times$ 0.1 mag).
If we had adopted the correction, it would have decreased the distance 
modulus only by 0.02 mag, or 1\% in distance.
The uncertainty in the metallicity dependence ($\pm$0.16 mag dex$^{-1} 
\times$ 0.15 dex) is treated as a random error.

\section{Barred Galaxies as Tully--Fisher Calibrators}

NGC~3319 was selected as a target galaxy by the H$_0$ Key Project
because it is a Tully--Fisher calibrator.
A full analysis of the calibration of the Tully--Fisher relation 
and its application to measure the H$_0$ is explored by Sakai et al.(1999).
However, in this particular paper, we are interested in 
the question of whether or not barred galaxies are
indeed ``ideal'' calibrators.
In Figure~7, we show on top the I--band Tully--Fisher relation
for 19 local galaxies whose distances have been measured accurately
from the Cepheid observations.  The total I--band magnitudes (Macri et
al. 1999) corrected for Galactic (Schlegel et al.) and internal extinction 
(Tully et al. 1997) are plotted against the 20\% linewidth measurements.
The inclination angles are those determined from surface photometry.
The overplotted line is the  bivariate least--squares fit:
$ M_I = -21.07 - 9.40 (\log W - 2.5)$.
The dispersion of this relation is 0.38 mag.
In the middle figure, we have used instead the kinematical inclination
angles for the six barred galaxies (NGC~925, 1365, 3319, 3351, 4535 and
4725).
The bivariate fit is now $M_I = -21.02 - 8.62 (\log W - 2.5)$, with
the dispersion of 0.32 mag, slightly smaller than the previous case.
Although the slope has changed by $\sim$0.8, as will be shown fully in
Sakai et al. (1999), this has little effect on the value of H$_0$ which 
is the main objective of the H$_0$ Key Project.
However, the change in the zero point affects the H$_0$ more directly.
In the bottom figure, the kinematical inclination angles were
used wherever available, whether the galaxy is barred or not.
The bivariate fit to this third set of Tully--Fisher data gives
$M_I = -21.02 - 8.96 (\log W - 2.5)$.

The differences bewtween these fits suggest that the difference between
photometric and kinematical inclination angles for barred galaxies
are indeed large, significant enough to make a Tully--Fisher zero point
difference of few percent.  For barred galaxies, it is likely that
the photometric inclination angles are over--estimated.  Using the
kinematical inclination angles moves these galaxies to larger
linewidth on the Tully--Fisher diagrams, yielding a smaller dispersion.
This subject will be again discussed more rigorously in Sakai et al.
(1999), especially how this bias would affect the value of H$_0$.

\section{Conclusion}

Using WFPC2 on HST, we have observed NGC~3319 at thirteen epochs over
a two--year period, twelve observations of which were made within a 49--day window
in late 1997.  
From these data, we have discovered \numceph Cepheid variable stars.
A sample of 28 of these were used to estimate the distance to NGC~3319.
Adopting a true distance modulus and reddening for the LMC of
$18.50 \pm 0.10$ mag and $E(B-V) = 0.10$ mag, 
we find a true distance modulus to NGC~3319 of 
\distmod0 $\pm$ \distmoderrorrandom (random) $\pm$
\distmoderrorsys (systematic), corresponding to a distance of
14.3 $\pm$ 0.9 (random) $\pm$ 0.7 (systematic) Mpc.

Although NGC~3319 is a relatively isolated galaxy, with 
its nearest neighbor 4$^{\circ}$ away in the plane of the sky,
it is cataloged to be in a small group of four galaxies
(Tully 1988) :  NGC~3198, an SBc galaxy,
NGC~3184, an Scd galaxy, 
NGC~3104, an irregular galaxy, and NGC~3319.
NGC~3198 has also been observed by the H$_0$ Key Project and its distance
was determined to be $\mu_0 = 30.80 \pm 0.16$ (random) $\pm$ 0.12 (systematic)
(Kelson et al. 1999).
The Cepheid distance reported in this paper is in excellent agreement
with that of NGC~3198.

\bigskip

The work presented in this paper is based on observations with the
NASA/ESA Hubble Space Telescope, obtained by the Space Telescope 
Science Institute, which is operated by AURA, Inc. under NASA contract 
No. 5-26555. We gratefully acknowledge the support of the NASA and STScI 
support staff, with special thanks to  our program coordinator, Doug Van 
Orsow. Support for this work was  provided by NASA through grant GO-2227-87A 
from STScI.
SS would like to acknowledge the support by NASA through the 
Long Term Space Astrophysics Program (NAS 7$-$1260).
LF acknowledges support by NASA through Hubble Fellowship grant
HF$-$01081.01$-$96A   awarded by the Space Telescope Science Institute.
SMGH and PBS acknowledge support from a NATO collaborative
research grant (CRG960178).
This research has made use of the NASA/IPAC Extragalactic Database (NED) 
which is operated by the Jet Propulsion Laboratory, Caltech, under contract 
with the National Aeronautics and Space Administration.

\begin{deluxetable}{lccccc}
\tablecolumns{6}
\tablewidth{0pc}
\small
\tablecaption{\bf HST Observations of NGC~3319}
\tablehead{
\colhead{Epoch} &
\colhead{Filename} &
\colhead{Date} &
\colhead{Julian Date} &
\colhead{Exposure Time} &
\colhead{Filter}
}
\startdata
1  & u2S75501t/2t & 1996 Jan 01 &  2450083.905  & 1100  $\times$ 2 & F555W        \nl
2  & u34l5101r/2r & 1997 Nov 17 &  2450769.984  & 1100  $\times$ 2 & F555W \nl
3  & u34l5201r/2r & 1997 Nov 24 &  2450776.905  & 1100  $\times$ 2 & F555W      \nl
4  & u34l5301r/2r & 1997 Dec 02 &  2450784.697  & 1100  $\times$ 2 & F555W      \nl
5  & u34l5401r/2r & 1997 Dec 04 &  2450786.108  & 1100  $\times$ 2 & F555W      \nl 
6  & u34l5501r/2r & 1997 Dec 06 &  2450788.662  & 1100  $\times$ 2 & F555W      \nl
7  & u34l5601r    & 1997 Dec 08 &  2450790.884  & 2200  $\times$ 1 & F555W      \nl
8  & u34l5701r/2r & 1997 Dec 11 &  2450793.701  & 1100  $\times$ 2 & F555W      \nl
9  & u34l5801r/2r & 1997 Dec 14 &  2450796.926  & 1100  $\times$ 2 & F555W      \nl
10 & u34l5901r/2r & 1997 Dec 18 &  2450800.958  & 1100  $\times$ 2 & F555W        \nl
11 & u34l6001r/2r & 1997 Dec 22 &  2450804.990  & 1100  $\times$ 2 & F555W        \nl
12 & u34l6101r/2r & 1997 Dec 27 &  2450809.845  & 1100  $\times$ 2 & F555W        \nl
13 & u34l6201r/2r & 1998 Jan 03 &  2450816.832  & 1100  $\times$ 2 & F555W        \nl
   &    &  &  &  &  \nl
2  & u34l5103r/4r & 1997 Nov 17 &  2450770.049  & 1300  $\times$ 2 & F814W    \nl  
3  & u34l5203r/4r & 1997 Nov 24 &  2450776.971  & 1300  $\times$ 2 & F814W      \nl
7  & u34l5602r/3r & 1997 Dec 08 &  2450790.948  & 1300  $\times$ 2 & F814W      \nl
12 & u34l6103r/4r & 1997 Dec 27 &  2450809.902  & 1300  $\times$ 2 & F814W        \nl
\enddata
\label{table:obslog}
\end{deluxetable}

\begin{deluxetable}{ccccc}
\tablecolumns{5}
\tablewidth{0pc}
\small
\tablecaption{\bf Photometric transformation constants for ALLFRAME photometry}
\tablehead{
\colhead{Filter} &
\colhead{Chip}&
\colhead{C1}&
\colhead{C2}&
\colhead{C3}
}
\startdata
F555W & 1 & $-0.969$ & $-0.052$ & $0.027$ \\
& 2 & $-0.957$ & $-0.052$ & $0.027$ \\
& 3 & $-0.949$ & $-0.052$ & $0.027$ \\
& 4 & $-0.973$ & $-0.052$ & $0.027$ \\
  & & & \\
F814W & 1 & $-1.863$ & $-0.063$ & $0.025$ \\
& 2 & $-1.822$ & $-0.063$ & $0.025$ \\
& 3 & $-1.841$ & $-0.063$ & $0.025$ \\
& 4 & $-1.870$ & $-0.063$ & $0.025$ \\
  & & & \\
\enddata
\end{deluxetable}

\begin{deluxetable}{lccccccc}
\tablecolumns{8}
\tablewidth{0pc}
\small
\tablecaption{\bf Reference Star Photometry}
\tablehead{
\colhead{ID} &
\colhead{Chip} &
\colhead{X} &
\colhead{Y} &
\colhead{RA} &
\colhead{DEC} &
\colhead{V (mag)} &
\colhead{I (mag)} \\
\colhead{} &
\colhead{} &
\colhead{pix} &
\colhead{pix} &
\colhead{J2000} &
\colhead{J2000} &
\colhead{mag} &
\colhead{mag} 
}

\startdata
  R01 &   1  &  460.8 &  165.3 &  10:39:09.31  &  41:42:09.6  & 23.00 $\pm$  0.01 &  22.94 $\pm$  0.01  \nl
  R02 &   1  &  450.5 &  552.9 &  10:39:08.62  &  41:42:25.4  & 21.08 $\pm$  0.01 &  20.79 $\pm$  0.01  \nl
  R03 &   1  &  374.5 &  170.9 &  10:39:09.61  &  41:42:11.7  & 24.15 $\pm$  0.01 &  24.22 $\pm$  0.02  \nl
  R04 &   1  &  176.9 &  588.3 &  10:39:09.54  &  41:42:32.6  & 23.74 $\pm$  0.01 &  23.91 $\pm$  0.03  \nl
  R05 &   1  &  288.3 &  426.7 &  10:39:09.44  &  41:42:23.8  & 23.77 $\pm$  0.01 &  23.79 $\pm$  0.02  \nl
  R06 &   1  &  518.7 &  303.0 &  10:39:08.84  &  41:42:13.9  & 24.55 $\pm$  0.01 &  22.65 $\pm$  0.01  \nl
  R07 &   1  &  363.6 &  564.8 &  10:39:08.91  &  41:42:27.7  & 24.69 $\pm$  0.01 &  24.57 $\pm$  0.05  \nl
  R08 &   1  &  342.0 &  676.9 &  10:39:08.78  &  41:42:32.7  & 24.78 $\pm$  0.01 &  24.78 $\pm$  0.05  \nl
  R09 &   1  &  168.8 &  481.4 &  10:39:09.77  &  41:42:28.5  & 25.58 $\pm$  0.01 &  25.72 $\pm$  0.08  \nl
  R10 &   1  &  399.6 &  168.9 &  10:39:09.53  &  41:42:11.1  & 25.36 $\pm$  0.02 &  25.51 $\pm$  0.06  \nl
  R11 &   1  &  550.6 &  485.3 &  10:39:08.39  &  41:42:20.6  & 25.00 $\pm$  0.02 &  24.83 $\pm$  0.05  \nl
  R12 &   1  &  435.5 &  421.3 &  10:39:08.92  &  41:42:20.4  & 25.72 $\pm$  0.02 &  25.72 $\pm$  0.08  \nl
  R13 &   1  &  139.9 &  308.0 &  10:39:10.20  &  41:42:22.1  & 25.19 $\pm$  0.02 &  25.11 $\pm$  0.05  \nl
  R14 &   1  &  210.3 &  318.7 &  10:39:09.92  &  41:42:21.1  & 25.45 $\pm$  0.02 &  25.38 $\pm$  0.06  \nl
  R15 &   1  &  407.9 &  322.8 &  10:39:09.21  &  41:42:17.1  & 24.49 $\pm$  0.02 &  24.70 $\pm$  0.03  \nl
  R16 &   1  &  203.5 &  400.2 &  10:39:09.80  &  41:42:24.5  & 25.59 $\pm$  0.02 &  25.55 $\pm$  0.09  \nl
  R17 &   1  &  201.3 &  115.3 &  10:39:10.34  &  41:42:13.1  & 25.33 $\pm$  0.02 &  23.47 $\pm$  0.01  \nl
  R18 &   2  &  223.9 &  213.0 &  10:39:11.77   &   41:42:37.8  & 24.40 $\pm$  0.01 &  24.38 $\pm$  0.05 \nl
  R19 &   2  &  687.8 &  244.1 &  10:39:10.08   &   41:43:19.9  & 24.05 $\pm$  0.01 &  23.94 $\pm$  0.03  \nl
  R20 &   2  &  126.3 &  496.0 &  10:39:14.39   &   41:42:42.6  & 24.05 $\pm$  0.01 &  24.11 $\pm$  0.02  \\
  R21 &   2  &   93.3 &  341.3 &  10:39:13.32   &   41:42:32.4  & 24.16 $\pm$  0.01 &  24.02 $\pm$  0.02  \nl
  R22 &   2  &  372.0 &  432.8 &  10:39:12.87   &   41:43:01.1  & 24.34 $\pm$  0.01 &  23.88 $\pm$  0.02  \nl
  R23 &   2  &  304.3 &  485.3 &  10:39:13.56   &   41:42:57.6  & 24.83 $\pm$  0.01 &  24.73 $\pm$  0.03  \nl
  R24 &   2  &  554.6 &  214.5 &  10:39:10.40   &   41:43:06.9  & 25.25 $\pm$  0.01 &  24.25 $\pm$  0.03  \nl
  R25 &   2  &  686.3 &  329.5 &  10:39:10.75   &   41:43:23.8  & 24.15 $\pm$  0.01 &  23.58 $\pm$  0.02  \nl
  R26 &   2  &  245.6 &  545.9 &  10:39:14.28   &   41:42:55.3  & 24.76 $\pm$  0.01 &  22.23 $\pm$  0.01  \nl
  R27 &   2  &  697.9 &  701.3 &  10:39:13.60   &   41:43:42.0  & 24.99 $\pm$  0.01 &  25.11 $\pm$  0.06  \nl
  R28 &   2  &  656.9 &  298.4 &  10:39:10.63   &   41:43:19.7  & 24.74 $\pm$  0.02 &  24.05 $\pm$  0.03  \nl
  R29 &   2  &  657.1 &  451.8 &  10:39:11.83   &   41:43:26.9  & 25.02 $\pm$  0.02 &  23.27 $\pm$  0.03  \nl
  R30 &   2  &  286.2 &  617.9 &  10:39:14.68   &   41:43:02.3  & 25.33 $\pm$  0.02 &  25.02 $\pm$  0.08  \nl
  R31 &   2  &  718.0 &  154.7 &  10:39:09.26   &   41:43:18.3  & 25.21 $\pm$  0.02 &  25.24 $\pm$  0.05  \nl
  R32 &   2  &  608.6 &  150.9 &  10:39:09.68   &   41:43:08.6  & 25.38 $\pm$  0.02 &  25.25 $\pm$  0.07  \nl
  R33 &   3  &  413.4 &  142.0 & 10:39:14.41   & 41:42:23.1   & 25.92 $\pm$  0.03 &  24.19 $\pm$  0.02  \nl
  R34 &   3  &  414.7 &  429.5 & 10:39:15.61   & 41:41:57.9   & 25.39 $\pm$  0.02 &  25.15 $\pm$  0.04  \nl
  R35 &   3  &  400.3 &  138.1 & 10:39:14.29   & 41:42:22.9   & 24.19 $\pm$  0.01 &  24.15 $\pm$  0.02  \nl
  R36 &   3  &  373.6 &  445.3 & 10:39:15.35   & 41:41:54.6   & 25.45 $\pm$  0.02 &  25.33 $\pm$  0.05  \nl
  R37 &   3  &  707.1 &  452.7 & 10:39:18.00   & 41:42:09.4   & 25.09 $\pm$  0.02 &  25.03 $\pm$  0.06  \nl
  R38 &   3  &  464.5 &  604.3 & 10:39:16.72   & 41:41:44.9   & 25.59 $\pm$  0.02 &  24.25 $\pm$  0.03  \nl
  R39 &   3  &  581.4 &  309.9 & 10:39:16.42   & 41:42:16.2   & 25.27 $\pm$  0.02 &  25.47 $\pm$  0.04  \nl
  R40 &   3  &  444.2 &  269.6 & 10:39:15.18   & 41:42:13.4   & 24.54 $\pm$  0.01 &  24.32 $\pm$  0.02  \nl
  R41 &   3  &  540.5 &  225.1 & 10:39:15.75   & 41:42:21.7   & 24.77 $\pm$  0.01 &  24.75 $\pm$  0.03  \nl
  R42 &   3  &  497.4 &  279.5 & 10:39:15.64   & 41:42:15.0   & 25.11 $\pm$  0.01 &  23.11 $\pm$  0.01  \nl
  R43 &   3  &  665.9 &  304.0 & 10:39:17.06   & 41:42:20.6   & 25.48 $\pm$  0.03 &  25.41 $\pm$  0.06  \nl
  R44 &   3  &  603.2 &  267.7 & 10:39:16.42   & 41:42:20.9   & 25.07 $\pm$  0.01 &  25.16 $\pm$  0.07  \nl
  R45 &   3  &  745.3 &  393.4 & 10:39:18.05   & 41:42:16.4   & 25.06 $\pm$  0.02 &  25.25 $\pm$  0.05  \nl
  R46 &   3  &  326.6 &  141.9 & 10:39:13.73   & 41:42:19.1   & 24.72 $\pm$  0.01 &  24.47 $\pm$  0.05  \nl
  R47 &   3  &  363.7 &  246.3 & 10:39:14.45   & 41:42:11.7   & 25.07 $\pm$  0.01 &  25.24 $\pm$  0.05  \nl
  R48 &   3  &  237.4 &  704.3 & 10:39:15.35   & 41:41:25.7   & 25.42 $\pm$  0.03 &  23.83 $\pm$  0.02  \nl
  R49 &   3  &  351.5 &  519.6 & 10:39:15.48   & 41:41:47.1   & 25.40 $\pm$  0.02 &  25.50 $\pm$  0.05  \nl
  R50 &   3  &  264.0 &  567.8 & 10:39:15.00   & 41:41:38.8   & 25.60 $\pm$  0.02 &  23.73 $\pm$  0.02  \nl
  R51 &   3  &  667.9 &  511.9 & 10:39:17.93   & 41:42:02.4   & 25.00 $\pm$  0.02 &  23.93 $\pm$  0.03  \nl
  R52 &   3  &  517.0 &  693.5 & 10:39:17.50   & 41:41:39.5   & 25.98 $\pm$  0.03 &  24.58 $\pm$  0.02  \nl
  R53 &   3  &  355.7 &  488.9 & 10:39:15.39   & 41:41:50.0   & 24.97 $\pm$  0.01 &  23.28 $\pm$  0.02  \nl
  R54 &   3  &  558.4 &  705.1 & 10:39:17.87   & 41:41:40.5   & 25.77 $\pm$  0.02 &  24.49 $\pm$  0.03  \nl
  R55 &   4  &  424.0 &  174.5 & 10:39:11.57   &  41:41:34.6   & 23.01 $\pm$  0.01 &  22.40 $\pm$  0.03  \nl
  R56 &   4  &  516.1 &  169.1 & 10:39:11.98   &  41:41:26.7   & 23.13 $\pm$  0.01 &  22.07 $\pm$  0.02  \nl
  R57 &   4  &  654.9 &  271.9 & 10:39:11.73   &  41:41:09.8   & 24.63 $\pm$  0.03 &  22.56 $\pm$  0.03  \nl
  R58 &   4  &  178.4 &   94.9 & 10:39:11.19   &  41:41:59.8   & 23.24 $\pm$  0.02 &  23.24 $\pm$  0.04  \nl
  R59 &   4  &  722.1 &  264.7 & 10:39:12.06   &  41:41:04.2   & 24.03 $\pm$  0.02 &  22.87 $\pm$  0.03  \nl
  R60 &   4  &  618.2 &  134.4 & 10:39:12.66   &  41:41:19.3   & 21.94 $\pm$  0.00 &  21.55 $\pm$  0.02  \nl
  R61 &   4  &  337.1 &  393.1 & 10:39:09.49   &  41:41:32.3   & 23.33 $\pm$  0.01 &  22.66 $\pm$  0.02  \nl
  R62 &   4  &  420.8 &  584.6 & 10:39:08.32   &  41:41:16.2   & 23.54 $\pm$  0.02 &  22.46 $\pm$  0.04  \nl
  R63 &   4  &  116.9 &  671.4 & 10:39:06.41   &  41:41:39.0   & 25.93 $\pm$  0.03 &  24.10 $\pm$  0.03  \nl
  R64 &   4  &  664.7 &  616.1 & 10:39:09.06   &  41:40:53.3   & 23.70 $\pm$  0.01 &  23.59 $\pm$  0.05  \nl
  R65 &   4  &  744.4 &  587.8 & 10:39:09.60   &  41:40:47.6   & 24.73 $\pm$  0.01 &  24.29 $\pm$  0.03  \nl
  R66 &   4  &  146.2 &  529.0 & 10:39:07.64   &  41:41:43.0   & 24.81 $\pm$  0.02 &  24.62 $\pm$  0.06  \nl
  R67 &   4  &  163.0 &  467.8 & 10:39:08.19   &  41:41:44.3   & 24.82 $\pm$  0.02 &  24.72 $\pm$  0.06  \nl
  R68 &   4  &  168.0 &  386.2 & 10:39:08.86   &  41:41:47.6   & 25.69 $\pm$  0.03 &  23.88 $\pm$  0.03  \nl
  R69 &   4  &  707.9 &  382.2 & 10:39:11.08   &  41:41:00.1   & 25.24 $\pm$  0.02 &  24.83 $\pm$  0.04  \nl
  R70 &   4  &  619.0 &  353.3 & 10:39:10.95   &  41:41:09.2   & 25.44 $\pm$  0.02 &  25.15 $\pm$  0.07  \nl
  R71 &   4  &  599.3 &  256.4 & 10:39:11.63   &  41:41:15.4   & 25.18 $\pm$  0.02 &  25.06 $\pm$  0.05  \nl
  R72 &   4  &  283.8 &  340.1 & 10:39:09.69   &  41:41:39.4   & 25.03 $\pm$  0.02 &  25.23 $\pm$  0.08  
\enddata			 	       
\label{table:refstars}
\end{deluxetable}

\begin{deluxetable}{ccrrcc}
\tablecolumns{6}
\tablewidth{0pc}
\small
\tablecaption{\bf Cepheid Positions} 
\tablehead{
\colhead{Cepheid} &
\colhead{Chip} &
\colhead{x\tablenotemark{a}} &
\colhead{y\tablenotemark{a}} &
\colhead{RA\tablenotemark{b}} &
\colhead{DEC\tablenotemark{b}} 
}
\startdata
 C01  &   4 &   77.4   &  480.6   &   10:39:07.75  & 41:41:51.2    \nl 
 C02  &   4 &  649.6   &  671.6   &   10:39:08.56  & 41:40:52.1    \nl
 C03  &   4 &  156.5   &  133.0   &   10:39:10.81  & 41:42:00.0    \nl
 C04  &   4 &  577.6   &  369.8   &   10:39:10.65  & 41:41:12.1    \nl
 C05  &   1 &  498.8   &  755.2   &   10:39:08.07  & 41:42:32.5    \nl
 C06  &   2 &  397.8   &  428.5   &   10:39:12.73  & 41:43:03.2    \nl
 C07  &   4 &  214.3   &  173.0   &   10:39:10.73  & 41:41:53.1    \nl
 C08  &   4 &  609.2   &  595.5   &   10:39:09.00  & 41:40:59.1    \nl
 C09  &   4 &  343.6   &  605.4   &   10:39:07.84  & 41:41:22.1    \nl
 C10  &   4 &  300.0   &  660.8   &   10:39:07.23  & 41:41:23.4    \nl
 C11  &   4 &  706.1   &  537.2   &   10:39:09.85  & 41:40:53.2    \nl
 C12  &   3 &   64.4   &   61.9   &   10:39:11.36  & 41:42:13.9    \nl
 C13  &   4 &  178.5   &  642.0   &   10:39:06.88  & 41:41:35.0    \nl
 C14  &   1 &  218.1   &  557.6   &   10:39:09.45  & 41:42:30.5    \nl
 C15  &   2 &  451.4   &  307.7   &   10:39:11.56  & 41:43:02.2    \nl
 C16  &   2 &  126.4   &  232.0   &   10:39:12.32  & 41:42:30.2    \nl
 C17  &   2 &  206.8   &   80.8   &   10:39:10.81  & 41:42:30.2    \nl
 C18  &   1 &  532.6   &  244.3   &   10:39:08.91  & 41:42:11.3    \nl
 C19  &   4 &   71.6   &  728.1   &   10:39:05.79  & 41:41:40.4    \nl
 C20  &   2 &  183.5   &  510.8   &   10:39:14.27  & 41:42:48.2    \nl
 C21  &   3 &  690.6   &  111.7   &   10:39:16.46  & 41:42:38.5    \nl
 C22  &   3 &  319.9   &  429.6   &   10:39:14.86  & 41:41:53.5    \nl
 C23  &   2 &  361.1   &  563.0   &   10:39:13.94  & 41:43:06.2    \nl
 C24  &   1 &  453.8   &  280.5   &   10:39:09.12  & 41:42:14.4    \nl
 C25  &   4 &  679.9   &  467.9   &   10:39:10.29  & 41:40:58.7    \nl
 C26  &   2 &   92.1   &  480.7   &   10:39:14.41  & 41:42:38.9    \nl
 C27  &   3 &  472.2   &  488.8   &   10:39:16.31  & 41:41:55.4    \nl
 C28  &   1 &  388.3   &  387.2   &   10:39:09.16  & 41:42:20.1    \nl
 C29  &   3 &  689.4   &  290.2   &   10:39:17.19  & 41:42:22.9    \nl
 C30  &   3 &  124.0   &  470.0   &   10:39:13.50  & 41:41:40.9    \nl
 C31  &   1 &  169.9   &  295.8   &   10:39:10.11  & 41:42:21.0    \nl
 C32  &   2 &   99.9   &  494.5   &   10:39:14.49  & 41:42:40.2    \nl
 C33  &   4 &  353.6   &  177.3   &   10:39:11.26  & 41:41:40.7    \nl
\enddata
\tablenotetext{a}{The x and y coordinates refer to the positions
of stars on the frame u34l5501r (second epoch of observations).}
\tablenotetext{b}{J2000}
\label{table:obslog}
\end{deluxetable}

\begin{deluxetable}{rcccccc}
\tablecolumns{7}
\tablewidth{0pc}
\small
\tablecaption{\bf V Magnitudes and Standard Errors of NGC~3319 Cepheid}
\tablehead{
\colhead{HJD} &
\colhead{V (mag)} &
\colhead{V (mag)} &
\colhead{V (mag)} &
\colhead{V (mag)} &
\colhead{V (mag)} &
\colhead{V (mag)} 
}
\startdata
 2450000+ &   C01 &   C02 &   C03 &   C04 &   C05 &   C06 \nl
    83.39  &   \nodata &  25.24 (0.22) &  25.67 (0.09) &  25.66 (0.17) &  \nodata &  26.44 (0.18) \nl 
    83.40  &   \nodata &  25.12 (0.18) &  25.50 (0.11) &  25.69 (0.13) &  \nodata &  26.18 (0.16) \nl
   769.47  &   24.89 (0.21) &  25.50 (0.17) &  25.46 (0.20) &  25.50 (0.18) &  25.07 (0.12) &  25.86 (0.16) \nl 
   769.48  &   24.97 (0.14) &  25.52 (0.19) &  25.44 (0.20) &  25.48 (0.14) &  25.24 (0.13) &  25.94 (0.16) \nl 
   776.39  &   24.99 (0.10) &  24.79 (0.09) &  25.55 (0.27) &  25.79 (0.17) &  25.35 (0.08) &  25.10 (0.41) \nl 
   776.41  &   25.06 (0.13) &  24.72 (0.11) &  24.03 (0.84) &  25.86 (0.22) &  25.38 (0.09) &  24.86 (0.25) \nl 
   784.18  &   24.80 (0.10) &  24.77 (0.17) &  25.45 (0.17) &  26.33 (0.22) &  26.74 (0.75) &  25.69 (0.09) \nl 
   784.20  &   24.89 (0.14) &  24.68 (0.16) &  26.82 (1.30) &  25.89 (0.24) &  25.43 (0.14) &  25.32 (0.74) \nl 
   785.59  &   24.65 (0.07) &  24.74 (0.14) &  25.31 (0.14) &  26.26 (0.23) &  25.02 (0.14) &  25.84 (0.12) \nl 
   785.61  &   24.82 (0.10) &  24.82 (0.12) &  25.28 (0.07) &  26.02 (0.18) &  24.95 (0.08) &  25.30 (0.71) \nl 
   788.15  &   24.44 (0.08) &  24.82 (0.10) &  25.04 (0.08) &  26.04 (0.20) &  24.64 (0.05) &  25.48 (0.10) \nl 
   788.16  &   24.47 (0.05) &  24.82 (0.10) &  23.56 (0.32) &  26.20 (0.23) &  24.61 (0.04) &  25.78 (0.15) \nl 
   790.37  &   24.31 (0.23) &  24.88 (0.10) &  25.30 (0.21) &  26.13 (0.15) &  24.59 (0.06) &  25.76 (0.05) \nl 
   790.37  &   24.31 (0.23) &  24.88 (0.10) &  25.30 (0.21) &  26.13 (0.15) &  24.59 (0.06) &  25.76 (0.05) \nl 
   793.19  &   24.20 (0.06) &  24.90 (0.06) &  25.03 (0.13) &  26.55 (0.72) &  24.59 (0.04) &  25.98 (0.14) \nl 
   793.20  &   24.30 (0.11) &  25.03 (0.13) &  25.05 (0.08) &  26.31 (0.52) &  24.60 (0.08) &  25.68 (0.12) \nl 
   796.41  &   24.17 (0.06) &  24.96 (0.12) &  25.22 (0.11) &  25.43 (0.17) &  24.82 (0.08) &  26.11 (0.09) \nl 
   796.43  &   24.29 (0.06) &  25.05 (0.09) &  25.11 (0.13) &  25.43 (0.18) &  24.83 (0.06) &  26.44 (0.15) \nl 
   800.44  &   24.31 (0.06) &  25.15 (0.09) &  25.12 (0.12) &  25.15 (0.10) &  24.86 (0.12) &  25.71 (0.09) \nl 
   800.46  &   24.35 (0.07) &  24.12 (0.56) &  25.16 (0.09) &  25.14 (0.12) &  24.93 (0.07) &  26.10 (0.18) \nl 
   804.48  &   24.50 (0.10) &  25.26 (0.11) &  25.13 (0.12) &  25.30 (0.17) &  23.16 (0.37) &  24.89 (0.07) \nl 
   804.49  &   24.61 (0.16) &  25.33 (0.20) &  25.19 (0.12) &  25.32 (0.13) &  25.11 (0.07) &  24.90 (0.08) \nl 
   809.33  &   24.47 (0.37) &  25.13 (0.13) &  25.31 (0.12) &  25.47 (0.16) &  25.09 (0.08) &  25.09 (0.10) \nl 
   809.35  &   24.54 (0.09) &  25.13 (0.18) &  25.42 (0.17) &  25.47 (0.19) &  25.16 (0.08) &  25.25 (0.12) \nl 
   816.32  &   24.83 (0.11) &  25.24 (0.47) &  25.67 (0.22) &  25.89 (0.14) &  25.47 (0.22) &  25.61 (0.24) \nl 
   816.33  &   24.67 (0.17) &  25.48 (0.18) &  25.59 (0.20) &  25.89 (0.18) &  25.43 (0.08) &  25.46 (0.10) \nl 
 2450000+ &   C07 &   C08 &   C09 &   C10 &   C11 &   C12 \nl
    83.39  &   25.35 (0.92) &  24.40 (0.06) &  25.34 (0.10) &  25.96 (0.22) &  25.13 (0.10) &  \nodata \nl 
    83.40  &   26.21 (0.21) &  24.66 (0.14) &  25.27 (0.14) &  25.88 (0.26) &  25.37 (0.17) &  \nodata \nl 
   769.47  &   25.97 (0.23) &  24.96 (0.31) &  25.62 (0.15) &  25.86 (0.18) &  25.34 (0.18) &  24.86 (0.07) \nl 
   769.48  &   25.89 (0.20) &  25.01 (0.12) &  25.87 (0.25) &  25.67 (0.17) &  25.22 (0.11) &  24.88 (0.09) \nl 
   776.39  &   25.21 (0.12) &  25.32 (0.14) &  25.82 (0.20) &  26.03 (0.26) &  25.36 (0.13) &  24.90 (0.19) \nl 
   776.41  &   25.50 (0.12) &  25.26 (0.23) &  26.11 (0.26) &  25.36 (0.22) &  25.52 (0.17) &  25.05 (0.12) \nl 
   784.18  &   25.75 (0.43) &  24.69 (0.12) &  26.14 (0.28) &  25.07 (0.10) &  24.63 (0.06) &  25.39 (0.11) \nl 
   784.20  &   26.02 (0.36) &  24.85 (0.23) &  26.18 (0.30) &  25.10 (0.18) &  24.43 (0.52) &  25.22 (0.10) \nl 
   785.59  &   25.77 (0.23) &  24.34 (0.09) &  25.46 (0.10) &  25.41 (0.18) &  24.78 (0.10) &  25.52 (0.07) \nl 
   785.61  &   25.86 (0.17) &  23.97 (0.20) &  25.65 (0.14) &  25.33 (0.11) &  24.74 (0.11) &  25.34 (0.12) \nl 
   788.15  &   25.82 (0.13) &  24.57 (0.11) &  24.97 (0.08) &  25.56 (0.19) &  24.83 (0.07) &  25.60 (0.14) \nl 
   788.16  &   26.14 (0.20) &  24.69 (0.09) &  25.03 (0.09) &  25.32 (0.12) &  25.05 (0.10) &  25.42 (0.13) \nl 
   790.37  &   26.18 (0.20) &  24.72 (0.21) &  25.20 (0.11) &  25.49 (0.09) &  25.13 (0.13) &  25.41 (0.10) \nl 
   790.37  &   26.18 (0.20) &  24.72 (0.21) &  25.20 (0.11) &  25.49 (0.09) &  25.13 (0.13) &  25.41 (0.10) \nl 
   793.19  &   26.01 (0.16) &  24.79 (0.09) &  25.11 (0.13) &  25.67 (0.16) &  25.22 (0.12) &  24.62 (0.06) \nl 
   793.20  &   25.97 (0.32) &  24.83 (0.11) &  25.41 (0.09) &  25.50 (0.17) &  25.25 (0.12) &  24.70 (0.08) \nl 
   796.41  &   26.00 (0.20) &  25.19 (0.11) &  25.39 (0.15) &  25.69 (0.24) &  25.29 (0.10) &  24.62 (0.06) \nl 
   796.43  &   26.01 (0.26) &  25.18 (0.14) &  25.65 (0.17) &  26.35 (0.34) &  25.33 (0.11) &  24.63 (0.10) \nl 
   800.44  &   25.33 (0.16) &  25.26 (0.13) &  25.73 (0.12) &  25.77 (0.17) &  25.36 (0.20) &  24.79 (0.10) \nl 
   800.46  &   25.66 (0.18) &  25.26 (0.14) &  25.72 (0.21) &  25.80 (0.18) &  25.48 (0.23) &  24.80 (0.09) \nl 
   804.48  &   25.24 (0.09) &  25.50 (0.18) &  25.82 (0.14) &  25.46 (0.26) &  25.21 (0.43) &  24.90 (0.08) \nl 
   804.49  &   25.23 (0.14) &  25.39 (0.13) &  25.71 (0.45) &  25.38 (0.29) &  25.91 (0.26) &  25.10 (0.14) \nl 
   809.33  &   25.69 (0.11) &  25.37 (0.20) &  26.25 (0.22) &  24.89 (0.09) &  24.76 (0.08) &  25.33 (0.11) \nl 
   809.35  &   25.95 (0.21) &  25.42 (0.23) &  25.83 (1.09) &  25.04 (0.08) &  24.80 (0.12) &  25.21 (0.10) \nl 
   816.32  &   25.94 (0.18) &  24.46 (0.16) &  24.98 (0.16) &  25.41 (0.20) &  25.00 (0.12) &  25.36 (0.11) \nl 
   816.33  &   26.11 (0.25) &  24.51 (0.07) &  24.92 (0.09) &  25.43 (0.17) &  24.97 (0.09) &  25.58 (0.13) \nl
 2450000+ &   C13 &   C14 &   C15 &   C16 &   C17 &   C18 \nl
    83.39  &   25.66 (0.11) &  23.81 (0.25) &  26.28 (0.15) &  25.43 (0.06) &  \nodata &  26.02 (0.14) \nl 
    83.40  &   26.09 (0.23) &  25.11 (0.04) &  26.30 (0.20) &  25.37 (0.14) &  \nodata &  25.61 (0.08) \nl 
   769.47  &   25.75 (0.20) &  26.10 (0.14) &  26.20 (0.36) &  25.83 (0.09) &  26.19 (0.12) &  25.56 (0.09) \nl 
   769.48  &   25.72 (0.14) &  25.67 (0.11) &  26.04 (0.10) &  25.82 (0.16) &  26.10 (0.13) &  25.63 (0.12) \nl 
   776.39  &   25.92 (0.23) &  24.96 (0.09) &  26.52 (0.17) &  25.54 (0.16) &  27.18 (0.67) &  26.18 (0.22) \nl 
   776.41  &   25.69 (0.14) &  24.99 (0.06) &  26.51 (0.21) &  25.48 (0.10) &  27.40 (1.63) &  26.25 (0.22) \nl 
   784.18  &   25.67 (0.19) &  25.74 (0.18) &  26.62 (0.29) &  26.00 (0.18) &  25.93 (0.18) &  26.04 (0.16) \nl 
   784.20  &   25.23 (0.20) &  25.37 (0.10) &  27.29 (0.31) &  26.03 (0.17) &  25.63 (0.11) &  26.07 (0.20) \nl 
   785.59  &   25.30 (0.17) &  25.66 (0.11) &  26.76 (0.23) &  26.02 (0.16) &  25.87 (0.13) &  25.30 (0.08) \nl 
   785.61  &   25.23 (0.11) &  25.41 (0.11) &  26.42 (0.14) &  26.07 (0.17) &  25.83 (0.16) &  25.20 (0.09) \nl 
   788.15  &   25.29 (0.13) &  25.78 (0.12) &  25.91 (0.09) &  26.39 (0.14) &  25.92 (0.17) &  25.37 (0.09) \nl 
   788.16  &   25.45 (0.15) &  25.79 (0.13) &  26.27 (0.14) &  26.08 (0.11) &  26.09 (0.22) &  25.55 (0.08) \nl 
   790.37  &   25.50 (0.11) &  25.72 (0.08) &  26.14 (0.14) &  26.30 (0.09) &  25.96 (0.09) &  25.68 (0.07) \nl 
   790.37  &   25.50 (0.11) &  25.72 (0.08) &  26.14 (0.14) &  26.30 (0.09) &  25.96 (0.09) &  25.68 (0.07) \nl 
   793.19  &   25.62 (0.17) &  25.79 (0.16) &  26.31 (0.15) &  25.72 (0.11) &  26.42 (0.19) &  26.07 (0.14) \nl 
   793.20  &   25.50 (0.14) &  25.73 (0.26) &  26.15 (0.13) &  25.50 (0.20) &  26.41 (0.24) &  25.99 (0.15) \nl 
   796.41  &   25.89 (0.21) &  24.89 (0.56) &  26.10 (0.12) &  25.14 (0.08) &  26.34 (0.19) &  24.10 (0.52) \nl 
   796.43  &   25.81 (0.22) &  25.37 (0.11) &  26.48 (0.20) &  25.16 (0.06) &  26.44 (0.17) &  26.42 (0.25) \nl 
   800.44  &   26.15 (0.23) &  25.12 (0.08) &  26.92 (0.26) &  25.56 (0.10) &  26.76 (0.21) &  26.21 (0.41) \nl 
   800.46  &   26.47 (0.46) &  25.08 (0.09) &  26.77 (0.29) &  25.60 (0.14) &  26.61 (0.18) &  26.17 (0.41) \nl 
   804.48  &   26.00 (0.25) &  25.25 (0.10) &  26.86 (0.23) &  25.85 (0.11) &  26.63 (0.22) &  26.20 (0.17) \nl 
   804.49  &   25.99 (0.26) &  24.17 (0.57) &  26.37 (0.64) &  25.78 (0.13) &  26.56 (0.25) &  26.42 (0.21) \nl 
   809.33  &   25.42 (0.13) &  25.54 (0.13) &  26.39 (0.12) &  26.04 (0.14) &  25.64 (0.19) &  25.55 (0.09) \nl 
   809.35  &   25.30 (0.16) &  25.57 (0.08) &  26.46 (0.15) &  26.02 (0.14) &  26.05 (0.34) &  25.53 (0.09) \nl 
   816.32  &   24.98 (0.59) &  25.25 (0.55) &  26.21 (0.23) &  25.47 (0.14) &  26.50 (0.18) &  25.85 (0.12) \nl 
   816.33  &   26.06 (0.87) &  25.77 (0.11) &  26.46 (0.19) &  25.14 (0.08) &  26.39 (0.17) &  26.16 (0.19) \nl 
 2450000+ &   C19 &   C20 &   C21 &   C22 &   C23 &   C24 \nl
    83.39  &   \nodata &  26.00 (0.22) &  25.22 (0.10) &  26.26 (0.17) &  25.73 (0.09) &  25.70 (0.13) \nl 
    83.40  &   \nodata &  26.21 (0.21) &  25.64 (0.36) &  26.60 (0.19) &  25.64 (0.10) &  25.80 (0.14) \nl 
   769.47  &   25.57 (0.16) &  26.25 (0.54) &  26.50 (0.21) &  25.87 (0.12) &  26.43 (0.12) &  26.03 (0.15) \nl 
   769.48  &   25.41 (0.14) &  26.38 (0.19) &  26.65 (0.25) &  25.99 (0.12) &  26.63 (0.27) &  25.98 (0.17) \nl 
   776.39  &   25.33 (0.11) &  25.95 (0.12) &  25.62 (0.13) &  26.25 (0.14) &  25.93 (0.11) &  26.78 (0.23) \nl 
   776.41  &   25.26 (0.11) &  25.86 (0.12) &  25.52 (0.12) &  26.49 (0.24) &  26.15 (0.44) &  26.58 (0.22) \nl 
   784.18  &   25.57 (0.15) &  26.35 (0.12) &  26.11 (0.17) &  25.90 (0.14) &  26.45 (0.20) &  25.96 (0.17) \nl 
   784.20  &   25.72 (0.32) &  26.36 (0.23) &  25.93 (0.19) &  25.91 (0.14) &  26.34 (0.20) &  25.95 (0.21) \nl 
   785.59  &   25.90 (0.38) &  26.16 (0.23) &  26.09 (0.18) &  26.26 (0.14) &  26.19 (0.13) &  25.91 (0.11) \nl 
   785.61  &   25.77 (0.19) &  26.11 (0.18) &  26.24 (0.15) &  25.83 (0.12) &  26.67 (0.19) &  26.09 (0.21) \nl 
   788.15  &   25.52 (0.10) &  26.49 (0.17) &  26.44 (0.23) &  26.18 (0.17) &  26.30 (0.16) &  26.04 (0.18) \nl 
   788.16  &   25.56 (0.14) &  26.93 (0.32) &  26.72 (0.27) &  26.24 (0.60) &  26.47 (0.16) &  26.07 (0.14) \nl 
   790.37  &   25.07 (0.12) &  26.73 (0.17) &  26.47 (0.12) & \nodata &  26.58 (0.21) & \nodata \nl 
   790.37  &   25.07 (0.12) &  26.73 (0.17) &  26.47 (0.12) & \nodata &  26.58 (0.21) & \nodata \nl 
   793.19  &   25.00 (0.17) &  25.57 (0.12) &  25.78 (0.12) &  26.39 (0.15) &  26.47 (0.22) &  26.45 (0.13) \nl 
   793.20  &   25.09 (0.10) &  25.58 (0.10) &  25.54 (0.38) &  27.43 (0.40) &  26.05 (0.18) &  26.51 (0.15) \nl 
   796.41  &   25.15 (0.51) &  25.83 (0.10) &  25.64 (0.11) &  26.79 (0.22) &  25.68 (0.08) &  26.49 (0.12) \nl 
   796.43  &   25.24 (0.15) &  25.89 (0.17) &  25.67 (0.08) &  26.07 (0.13) &  25.91 (0.09) &  26.82 (0.21) \nl 
   800.44  &   25.72 (0.21) &  26.17 (0.13) &  25.90 (0.12) &  25.55 (0.12) &  26.16 (0.12) &  26.13 (0.17) \nl 
   800.46  &   25.35 (0.22) &  25.94 (0.12) &  25.59 (0.29) &  25.63 (0.07) &  26.16 (0.08) &  25.54 (0.06) \nl 
   804.48  &   25.84 (0.17) &  26.24 (0.20) &  26.16 (0.16) &  25.88 (0.13) &  26.60 (0.25) &  26.01 (0.09) \nl 
   804.49  &   25.78 (0.20) &  26.41 (0.20) &  26.09 (0.13) &  26.52 (0.20) &  26.35 (0.14) &  26.14 (0.12) \nl 
   809.33  &   25.53 (0.12) &  26.65 (0.25) &  26.46 (0.19) &  26.90 (0.32) &  26.03 (0.15) &  26.02 (0.27) \nl 
   809.35  &   25.35 (0.48) &  26.48 (0.15) &  26.54 (0.28) &  26.76 (0.24) &  26.28 (0.17) &  27.09 (0.63) \nl 
   816.32  &   25.25 (0.16) &  25.83 (0.14) &  26.05 (0.20) &  23.97 (0.26) &  25.92 (0.12) &  26.03 (0.18) \nl 
   816.33  &   25.13 (0.14) &  26.00 (0.12) &  25.58 (0.17) &  25.35 (0.09) &  25.81 (0.12) &  25.82 (0.16) \nl
 2450000+ &   C25 &   C26 &   C27 &   C28 &   C29 &   C30 \nl
    83.39  &   26.37 (0.30) &  25.92 (0.08) &  27.89 (0.41) &  27.04 (0.45) &  26.03 (0.12) &  \nodata \nl 
    83.40  &   26.28 (0.16) &  26.11 (0.16) &  26.95 (0.29) &  26.85 (0.27) &  26.49 (0.28) &  \nodata \nl 
   769.47  &   26.31 (0.25) &  25.90 (0.10) &  26.37 (0.18) &  26.84 (0.30) &  25.92 (0.10) &  27.41 (0.60) \nl 
   769.48  &   26.60 (0.22) &  26.31 (0.26) &  26.38 (0.19) &  23.14 (0.44) &  26.21 (0.22) &  26.81 (0.26) \nl 
   776.39  &   25.88 (0.46) &  25.29 (0.10) &  27.01 (0.31) &  26.13 (0.13) &  25.62 (0.13) &  26.39 (0.31) \nl 
   776.41  &   25.76 (0.31) &  25.12 (0.11) &  27.02 (0.33) &  25.96 (0.14) &  25.66 (0.14) &  26.18 (0.32) \nl 
   784.18  &   26.10 (0.25) &  25.78 (0.09) &  26.06 (0.14) &  26.94 (0.30) &  26.21 (0.21) &  26.36 (0.23) \nl 
   784.20  &   25.99 (0.34) &  25.86 (0.11) &  26.34 (0.31) &  27.21 (0.37) &  26.31 (0.24) &  26.59 (0.34) \nl 
   785.59  &   26.37 (0.31) &  25.52 (0.89) &  26.40 (0.14) &  26.77 (0.24) &  26.04 (0.20) &  25.96 (0.18) \nl 
   785.61  &   26.37 (0.49) &  26.09 (0.19) &  26.15 (0.13) &  26.33 (0.19) &  26.00 (0.20) &  26.12 (0.15) \nl 
   788.15  &   26.36 (0.26) &  26.25 (0.12) &  24.40 (0.68) &  25.80 (0.10) &  25.71 (0.16) &  26.80 (0.21) \nl 
   788.16  &   26.34 (0.26) &  25.93 (0.13) &  26.75 (0.25) &  25.58 (0.09) &  25.35 (0.08) &  26.80 (0.27) \nl 
   790.37  &   26.32 (0.27) &  25.98 (0.09) &  26.77 (0.22) &  26.02 (0.09) &  25.71 (0.17) &  26.75 (0.29) \nl 
   790.37  &   26.32 (0.27) &  25.98 (0.09) &  26.77 (0.22) &  26.02 (0.09) &  25.71 (0.17) &  26.75 (0.29) \nl 
   793.19  &   25.84 (0.24) &  25.46 (0.10) &  26.90 (0.17) &  26.89 (0.32) &  26.00 (0.12) &  26.72 (0.28) \nl 
   793.20  &   25.98 (0.25) &  25.67 (0.19) &  26.91 (0.22) &  26.23 (0.18) &  25.99 (0.21) &  29.35 (3.26) \nl 
   796.41  &   25.94 (0.22) &  25.51 (0.08) &  26.44 (0.24) &  27.14 (0.33) &  26.01 (0.20) &  26.26 (0.35) \nl 
   796.43  &   25.80 (0.16) &  25.98 (0.14) &  26.86 (0.20) &  26.69 (0.23) &  26.24 (0.23) &  26.44 (0.17) \nl 
   800.44  &   26.33 (0.34) &  25.95 (0.15) &  26.04 (0.12) &  26.77 (0.24) &  25.53 (0.15) &  26.59 (0.29) \nl 
   800.46  &   26.08 (0.23) &  26.03 (0.11) &  26.04 (0.15) &  26.61 (0.17) &  25.58 (0.16) &  26.35 (0.18) \nl 
   804.48  &   26.71 (0.37) &  25.81 (0.18) &  26.67 (0.26) &  25.91 (0.15) &  25.85 (0.19) &  27.65 (0.71) \nl 
   804.49  &   26.24 (0.33) &  26.01 (0.24) &  26.91 (0.29) &  24.97 (0.63) &  25.99 (0.21) &  27.03 (0.32) \nl 
   809.33  &   25.97 (0.31) &  25.42 (0.05) &  26.98 (0.42) &  26.37 (0.14) &  26.17 (0.24) &  26.15 (0.14) \nl 
   809.35  &   26.08 (0.29) &  25.49 (0.11) &  26.97 (0.27) &  26.66 (0.21) &  26.23 (0.26) &  26.24 (0.22) \nl 
   816.32  &   26.19 (0.29) &  25.68 (0.08) &  26.15 (0.19) &  26.85 (0.19) &  25.84 (0.17) &  27.22 (0.30) \nl 
   816.33  &   25.96 (0.23) &  25.72 (0.13) &  26.54 (0.23) &  26.85 (0.26) &  25.83 (0.15) &  27.47 (0.41) \nl
 2450000+ &   C31 &   C32 &   C33 &    &   &   \nl
    83.39  &   26.23 (0.24) &  25.73 (0.13) &  26.10 (0.28) &    &    &    \nl 
    83.40  &   25.77 (0.10) &  26.28 (1.13) &  25.68 (0.31) &    &    &    \nl 
   769.47  &   26.11 (0.14) &  26.48 (0.21) &  25.70 (0.27) &    &    &    \nl 
   769.48  &   26.22 (0.24) &  26.48 (0.34) &  25.49 (0.32) &    &    &    \nl 
   776.39  &   26.67 (0.31) &  26.20 (0.17) &  26.37 (0.32) &    &    &    \nl 
   776.41  &   27.10 (0.38) &  26.30 (0.17) &  26.87 (0.92) &    &    &    \nl 
   784.18  &   25.88 (0.10) &  26.41 (0.31) &  26.41 (0.34) &    &    &    \nl 
   784.20  &   25.79 (0.12) &  26.55 (0.20) &  26.91 (0.68) &    &    &    \nl 
   785.59  &   26.24 (0.17) &  26.92 (0.25) &  25.72 (0.31) &    &    &    \nl 
   785.61  &   26.88 (0.36) &  27.37 (0.54) &  25.55 (0.24) &    &    &    \nl 
   788.15  &   26.67 (0.20) &  26.30 (0.22) &  25.82 (0.34) &    &    &    \nl 
   788.16  &   26.44 (0.17) &  26.09 (0.19) &  25.78 (0.18) &    &    &    \nl 
   790.37  &   26.99 (0.14) &  25.99 (0.14) &  25.90 (0.20) &    &    &    \nl 
   790.37  &   26.99 (0.14) &  25.99 (0.14) &  25.90 (0.20) &    &    &    \nl 
   793.19  &   26.38 (0.23) &  26.50 (0.23) &  26.12 (0.35) &    &    &    \nl 
   793.20  &   26.57 (0.26) &  26.76 (0.24) &  25.65 (0.29) &    &    &    \nl 
   796.41  &   25.84 (0.12) &  26.69 (0.20) &  25.99 (0.22) &    &    &    \nl 
   796.43  &   26.11 (0.12) &  26.54 (0.24) &  26.00 (0.30) &    &    &    \nl 
   800.44  &   26.92 (0.37) &  26.38 (0.19) &  26.16 (0.22) &    &    &    \nl 
   800.46  &   26.47 (0.20) &  25.99 (0.13) &  26.41 (0.48) &    &    &    \nl 
   804.48  &   26.96 (0.30) &  26.42 (0.22) &  26.07 (0.21) &    &    &    \nl 
   804.49  &   27.13 (0.26) &  26.51 (0.27) &  25.75 (0.29) &    &    &    \nl 
   809.33  &   26.13 (0.17) &  25.76 (1.32) &  26.05 (0.32) &    &    &    \nl 
   809.35  &   26.14 (0.17) &  26.87 (0.44) &  26.22 (0.36) &    &    &    \nl 
   816.32  &   28.37 (2.22) &  26.26 (0.20) &  26.21 (0.32) &    &    &    \nl 
   816.33  &   26.78 (0.23) &  26.38 (0.28) &  26.40 (0.27) &    &    &    \nl 
\enddata
\label{table:cephv}
\end{deluxetable}

\scriptsize

\def\Deg{\hbox{${}^\circ$\llap{.}}}
\def\Min{\hbox{${}^{\prime}$\llap{.}}}
\def\Sec{\hbox{${}^{\prime\prime}$\llap{.}}}

\def\numceph{33\thinspace}
\def\numcephall{49\thinspace}
\def\lowperiod{8\thinspace}
\def\upperperiod{47\thinspace}
\def\distmod0{30.78\thinspace}

\begin{deluxetable}{rcccccc}
\tablecolumns{7}
\tablewidth{0pc}
\small
\tablecaption{\bf  I Magnitudes and Standard Errors of NGC~3319 Cepheid}
\tablehead{
\colhead{HJD} &
\colhead{I (mag)} &
\colhead{I (mag)} &
\colhead{I (mag)} &
\colhead{I (mag)} &
\colhead{I (mag)} &
\colhead{I (mag)}
}
\startdata
 2450000+ &   C01 &   C02 &   C03 &   C04 &   C05 &   C06 \nl
   769.53  &   23.90 (0.22) &  24.27 (0.17) &  24.39 (0.11) &  24.17 (0.15) &  24.50 (0.07) &  24.73 (0.12) \nl 
   769.55  &   23.68 (0.36) &  24.11 (0.13) &  24.36 (0.14) &  24.16 (0.11) &  24.43 (0.06) &  24.65 (0.12) \nl 
   776.46  &   23.95 (0.09) &  23.66 (0.10) &  24.26 (0.21) &  24.66 (0.14) &  24.74 (0.14) &  24.16 (0.17) \nl 
   776.47  &   23.89 (0.10) &  23.77 (0.11) &  24.62 (0.27) &  24.72 (0.19) &  24.63 (0.10) &  24.43 (0.22) \nl 
   790.43  &   23.66 (0.17) &  23.51 (0.09) &  24.05 (0.12) &  24.75 (0.15) &  24.17 (0.06) &  24.79 (0.08) \nl 
   790.45  &   23.72 (0.13) &  23.64 (0.07) &  24.27 (0.10) &  24.90 (0.15) &  24.14 (0.09) &  24.65 (0.08) \nl 
   809.39  &   23.49 (0.06) &  24.04 (0.14) &  24.14 (0.08) &  24.25 (0.15) &  24.72 (0.13) &  24.27 (0.09) \nl 
   809.40  &   23.55 (0.19) &  24.01 (0.13) &  24.31 (0.12) &  24.12 (0.14) &  24.56 (0.10) &  24.61 (0.09) \nl 
  &  &  &  &  &  &  \nl
 2450000+ &   C07 &   C08 &   C09 &   C10 &   C11 &   C12 \nl
   769.53  &   24.95 (0.15) &  24.12 (0.16) &  24.57 (0.19) &  24.69 (0.14) &  24.48 (0.14) &  24.41 (0.15) \nl 
   769.55  &   25.01 (0.31) &  24.10 (0.22) &  24.62 (0.15) &  22.04 (0.44) &  24.42 (0.15) &  24.45 (0.15) \nl 
   776.46  &   24.28 (0.09) &  24.36 (0.19) &  24.71 (0.26) &  24.60 (0.44) &  24.38 (0.25) &  24.45 (0.16) \nl 
   776.47  &   22.23 (0.62) &  24.46 (0.21) &  24.60 (0.47) &  24.74 (0.22) &  24.61 (0.18) &  24.42 (0.13) \nl 
   790.43  &   24.96 (0.17) &  23.70 (0.12) &  24.34 (0.09) &  24.41 (0.11) &  24.49 (0.17) &  24.65 (0.09) \nl 
   790.45  &   24.92 (0.22) &  23.95 (0.14) &  24.28 (0.17) &  24.38 (0.15) &  24.49 (0.20) &  24.85 (0.12) \nl 
   809.39  &   24.48 (0.16) &  24.29 (0.12) &  25.05 (0.20) &  24.02 (0.13) &  24.30 (0.14) &  24.57 (0.09) \nl 
   809.40  &   24.49 (0.10) &  24.44 (0.18) &  24.88 (1.85) &  24.06 (0.07) &  24.40 (0.14) &  24.44 (0.34) \nl 
  &  &  &  &  &  &  \nl
 2450000+ &   C13 &   C14 &   C15 &   C16 &   C17 &   C18 \nl
   769.53  &   24.80 (0.09) &  24.95 (0.09) &  25.21 (0.12) &  25.11 (0.13) &  24.97 (0.09) &  24.84 (0.09) \nl 
   769.55  &   24.81 (0.20) &  25.06 (0.10) &  25.16 (0.17) &  25.15 (0.14) &  25.62 (0.21) &  24.94 (0.12) \nl 
   776.46  &   25.02 (0.29) &  24.69 (0.09) &  25.71 (0.20) &  24.54 (0.11) &  25.43 (0.30) &  25.44 (0.21) \nl 
   776.47  &   25.02 (0.20) &  24.43 (0.07) &  25.45 (0.19) &  24.54 (0.12) &  24.93 (0.18) &  25.23 (0.12) \nl 
   790.43  &   24.83 (0.20) &  25.01 (0.13) &  25.40 (0.13) &  25.12 (0.09) &  24.86 (0.11) &  24.88 (0.11) \nl 
   790.45  &   24.82 (0.21) &  25.01 (0.13) &  25.29 (0.16) &  25.20 (0.17) &  25.29 (0.44) &  24.93 (0.16) \nl 
   809.39  &   24.75 (0.20) &  24.49 (0.73) &  25.32 (0.20) &  24.84 (0.09) &  24.33 (0.21) &  24.89 (0.12) \nl 
   809.40  &   24.96 (0.22) &  24.48 (0.11) &  25.72 (0.13) &  25.12 (0.13) &  24.76 (0.12) &  24.79 (0.09) \nl
 2450000+ &   C19 &   C20 &   C21 &   C22 &   C23 &   C24 \nl
   769.53  &   24.94 (0.25) &  25.54 (0.20) &  25.38 (0.14) &  25.41 (0.17) &  24.98 (0.14) &  25.25 (0.18) \nl 
   769.55  &   24.70 (0.19) &  26.06 (0.40) &  26.10 (0.59) &  25.44 (0.18) &  25.15 (0.14) &  25.04 (0.12) \nl 
   776.46  &   24.75 (0.26) &  25.06 (0.16) &  24.83 (0.14) &  25.99 (0.23) &  24.84 (0.14) &  25.91 (0.29) \nl 
   776.47  &   24.64 (0.12) &  25.18 (0.18) &  24.93 (0.15) &  25.92 (0.18) &  24.68 (0.09) &  25.77 (0.21) \nl 
   790.43  &   24.55 (0.13) &  25.47 (1.34) &  25.27 (0.15) &  25.42 (0.21) &  25.07 (0.15) &  25.29 (0.09) \nl 
   790.45  &   24.54 (0.16) &  25.74 (0.14) &  25.55 (0.26) &  26.14 (0.30) &  24.79 (0.13) &  25.17 (0.16) \nl 
   809.39  &   24.79 (0.17) &  25.68 (0.18) &  25.47 (0.24) &  25.97 (0.20) &  25.01 (0.13) &  25.35 (0.18) \nl 
   809.40  &   24.77 (0.28) &  25.72 (0.22) &  25.95 (0.25) &  26.27 (0.33) &  24.90 (0.17) &  24.77 (0.75) \nl 
  &  &  &  &  &  &  \nl
 2450000+ &   C25 &   C26 &   C27 &   C28 &   C29 &   C30 \nl
   769.53  &   25.50 (0.29) &  24.50 (0.08) &  25.37 (0.17) &  25.62 (0.27) &  25.55 (0.20) &  27.12 (1.17) \nl 
   769.55  &   25.73 (0.28) &  24.87 (0.17) &  25.57 (0.18) &  25.53 (0.24) &  25.47 (0.24) &  25.94 (0.31) \nl 
   776.46  &   25.15 (0.66) &  24.42 (0.08) &  25.81 (0.12) &  25.30 (0.10) &  25.35 (0.35) &  25.72 (0.33) \nl 
   776.47  &   25.20 (0.53) &  24.43 (0.12) &  25.57 (0.17) &  25.07 (0.11) &  25.08 (0.13) &  26.27 (0.39) \nl 
   790.43  &   25.73 (0.62) &  24.61 (0.10) &  25.78 (0.17) &  25.10 (0.09) &  25.15 (0.15) &  26.07 (0.23) \nl 
   790.45  &   26.03 (0.78) &  24.75 (0.10) &  25.74 (0.19) &  24.98 (0.08) &  22.78 (0.47) &  25.92 (0.39) \nl 
   809.39  &   25.53 (0.27) &  24.47 (0.11) &  25.87 (0.25) &  25.37 (0.14) &  25.18 (0.17) &  23.57 (0.51) \nl 
   809.40  &   25.44 (0.36) &  24.35 (0.21) &  25.63 (0.17) &  25.23 (0.09) &  24.99 (0.15) &  25.78 (0.28) \nl 
  &  &  &  &  &  &  \nl
 2450000+ &   C31 &   C32 &   C33 &    &  &  \nl
   769.53  &   25.89 (0.19) &  25.63 (0.20) &  24.82 (0.28) &    &    &    \nl 
   769.55  &   26.74 (0.29) &  25.71 (0.28) &  24.68 (0.25) &    &    &    \nl 
   776.46  &   26.50 (0.50) &  25.59 (0.27) &  25.02 (0.35) &    &    &    \nl 
   776.47  &   25.85 (0.28) &  25.68 (0.39) &  24.94 (0.31) &    &    &    \nl 
   790.43  &   25.80 (0.17) &  25.78 (0.34) &  25.03 (0.36) &    &    &    \nl 
   790.45  &   26.71 (0.52) &  25.85 (0.38) &  25.07 (0.25) &    &    &    \nl 
   809.39  &   25.61 (0.27) &  25.14 (0.23) &  24.92 (0.32) &    &    &    \nl 
   809.40  &   25.69 (0.22) &  25.58 (0.27) &  25.98 (0.86) &    &    &    \nl 
\enddata
\label{table:cephmagi}
\end{deluxetable}

\begin{deluxetable}{cccccccrcl}
\tablecolumns{10}
\tablewidth{0pc}
\small
\tablecaption{\bf Cepheid Mean Magnitudes}
\tablehead{
\colhead{Cepheid} &
\colhead{Chip} &
\colhead{Use\tablenotemark{a}} &
\colhead{V$_{int}$} &
\colhead{V$_{ph}$} &
\colhead{I$_{int}$} &
\colhead{I$_{ph}$} &
\colhead{I$_{corr}$} &
\colhead{Period} &
\colhead{Comments}
}
\startdata
 C01  &   4 & n &    24.54 &  24.65 &  23.68 &  23.72  &  0.02 &   $>$47 &  \nl
 C02  &   4 & y &    25.02 &  25.03 &  23.84 &  23.83  &  0.02 &   49.3  &  \nl
 C03  &   4 & y &    25.30 &  25.31 &  24.23 &  24.22  & -0.04 &   43.6  &  \nl
 C04  &   4 & y &    25.67 &  25.65 &  24.41 &  24.45  & -0.08 &   39.5  &  \nl
 C05  &   1 & y &    24.97 &  24.97 &  24.43 &  24.41  &  0.02 &   33.4  &  \nl
 C06  &   2 & y &    25.55 &  25.48 &  24.62 &  24.58  &  0.01 &   31.0  &  \nl
 C07  &   4 & y &    25.78 &  25.74 &  24.63 &  24.63  & -0.03 &   28.4  &  \nl
 C08  &   4 & y &    24.83 &  24.94 &  24.02 &  24.06  & -0.05 &   28.1  &  \nl
 C09  &   4 & y &    25.49 &  25.59 &  24.54 &  24.59  & -0.03 &   28.1  &  \nl
 C10  &   4 & y &    25.47 &  25.43 &  24.46 &  24.43  &  0.01 &   27.5  &  \nl
 C11  &   4 & y &    25.10 &  25.14 &  24.45 &  24.47  &  0.01 &   26.3  &  \nl
 C12  &   3 & n &    25.06 &  25.01 &  24.50 &  24.48  & -0.05 &   26.0  & near the edge \nl
 C13  &   4 & y &    25.61 &  25.66 &  24.89 &  24.92  &  0.04 &   25.0  &  \nl
 C14  &   1 & n &    25.44 &  25.42 &  24.71 &  24.68  &  0.00 &   24.7  & blend \nl
 C15  &   2 & y &    26.36 &  26.40 &  25.46 &  25.50  &  0.04 &   23.4  &  \nl
 C16  &   2 & y &    25.69 &  25.71 &  24.82 &  24.78  & -0.04 &   22.9  &  \nl
 C17  &   2 & y &    26.11 &  26.20 &  25.11 &  25.17  &  0.17 &   22.5  &  \nl
 C18  &   1 & y &    25.79 &  25.89 &  24.99 &  25.06  &  0.04 &   21.2  &  \nl
 C19  &   4 & y &    25.40 &  25.38 &  24.73 &  24.71  &  0.03 &   20.1  &  \nl
 C20  &   2 & y &    26.13 &  26.07 &  25.39 &  25.31  & -0.07 &   19.9  &  \nl
 C21  &   3 & y &    25.94 &  25.93 &  25.16 &  25.10  & -0.02 &   19.8  &  \nl
 C22  &   3 & y &    26.01 &  25.98 &  25.52 &  25.54  & -0.11 &   18.1  &  \nl
 C23  &   2 & y &    26.15 &  26.13 &  24.86 &  24.86  & -0.04 &   17.8  &  \nl
 C24  &   1 & y &    26.04 &  26.10 &  25.38 &  25.44  & -0.06 &   17.5  &  \nl
 C25  &   4 & y &    26.13 &  26.10 &  25.51 &  25.49  & -0.08 &   17.1  &  \nl
 C26  &   2 & y &    25.75 &  25.74 &  24.61 &  24.59  &  0.06 &   16.7  &  \nl
 C27  &   3 & y &    26.52 &  26.61 &  25.65 &  25.71  & -0.02 &   16.1  &  \nl
 C28  &   1 & y &    26.36 &  26.32 &  25.28 &  25.26  & -0.04 &   14.9  &  \nl
 C29  &   3 & n &    25.87 &  25.86 &  25.20 &  25.18  & -0.01 &   12.5  & a bright nearby cluster \nl
 C30  &   3 & y &    26.53 &  26.56 &  25.97 &  25.98  &  0.07 &   12.4  &  \nl
 C31  &   1 & y &    26.30 &  26.35 &  25.74 &  25.74  & -0.01 &   12.1  &  \nl
 C32  &   2 & y &    26.31 &  26.32 &  25.45 &  25.46  & -0.02 &   12.0  &  \nl
 C33  &   4 & n &    25.89 &  25.90 &  24.94 &  24.95  &  0.05 &    8.2  &  \nl
\enddata
\label{table:cephmags}
\tablenotetext{a}{Whether that particular Cepheid variable was used the distance
determination or not.}
\end{deluxetable}

\begin{deluxetable}{lll}
\tablecolumns{3}
\tablewidth{0pc}
\scriptsize
\tablecaption{Error Budget in the Distance to NGC 3319\label{tbl-10}}
\tablehead{
\colhead{Source} &
\colhead{Error} &
\colhead{Notes} \nl
\colhead{ } &
\colhead{(mag)} &
\colhead{ } 
}
\startdata
{\bf I. CEPHEID PL CALIBRATION} & &  \nl
(A) LMC True Modulus & $\pm$ 0.10 &  \nl 
(B) VI Dereddened PL Zero Point\tablenotemark{a} & $\pm$ 0.02 &  \nl
(S1) PL Systematic Uncertainty&$\pm$ 0.10& (A) and (B) added in quadrature \nl
& & \nl
{\bf II. EXTINCTION CORRECTIONS} & &  \nl
(R1) Errors in the adopted value for R$_V$ & $\pm$ 0.01 & See Ferrarese et al. 1997 for details\nl 
& & \nl
{\bf III. NGC 3319 $-$ LMC METALLICITY DIFFERENCES} & &  \nl
(S2) Metallicity Differences& $\pm$ 0.02  & \nl
(R2) Uncertainty in the metallicity dependence & $\pm$ 0.02 & \nl
& & \nl
{\bf IV. NGC 3319 $-$ LMC R$_V$ DIFFERENCES} & &  \nl
(R3) R$_V$ Differences&$\pm$ 0.014 &See Ferrarese et al. 1997 for details\nl 
& & \nl
(
{\bf V. PHOTOMETRIC ERRORS} & &  \nl
{\it (D) HST V-Band Zero Point\tablenotemark{b}} & $\pm$ 0.07 &   \nl
{\it (E) HST I-Band Zero Point\tablenotemark{b}} & $\pm$ 0.07 &   \nl
(R4) Cepheid True Modulus\tablenotemark & $\pm$ 0.10 & (D) and (E) added in quadrature  \nl
& & \nl
{\bf VI. NGC 3319 PL FITTING} & &  \nl
{\it (F) PL Fit (V)} & $\pm$ 0.06 &  \nl
{\it (G) PL Fit (I)} & $\pm$ 0.06 &  \nl
(R5) Cepheid True Modulus\tablenotemark{c} &$\pm$ 0.09 &(F),(G) partially correlated.\nl
& & \nl
{\bf TOTAL UNCERTAINTY} & & \nl
(R) Random Errors&$\pm$ 0.14 &R1,R2,R3,R4,R5 added in quadrature \nl
(S) Systematic Errors&$\pm$ 0.10 &S1,S2 added in quadrature \nl
& & \nl
\enddata
\tablenotetext{a} {Derived from the observed scatter in the Madore \& Freedman (1991) 
PL relation, with 32  contributing LMC Cepheids.  Uncertainties in the V and I 
PL zero points are correlated.}
\tablenotetext{b} {Uncertainties from aperture corrections,
the Holtzman et al. (1995) zero points, the long-versus short uncertainty,
and the ALLFRAME aperture correction uncertainties, combined
in quadrature.}
\tablenotetext{c} {The partially correlated nature of the derived PL width uncertainties 
is taken into account by the (correlated) de$-$reddening procedure, coupled 
with the largely `degenerate-with-reddening' intrinsic positioning of individual 
Cepheids within the instability strip.}

\end{deluxetable}

\include{fig}

\end{document}